\documentclass[]{llncs}

\usepackage[latin1]{inputenc}
\usepackage[english]{babel}
\usepackage{amsmath}
\usepackage{amsfonts}
\usepackage{amssymb}
\usepackage{epsfig}
\usepackage{psfrag}
\usepackage{comment}
\usepackage{color}
\usepackage{psfrag}
\usepackage{soul}
\usepackage{subfigure}

\begin{document}

\title{Throughput Analysis in CSMA/CA Networks using Continuous Time Markov Networks: A Tutorial}

\author{B. Bellalta$^{1}$, A. Zocca$^{2}$, C. Cano$^{3}$, A. Checco$^{3}$, J. Barcelo$^{1}$, A. Vinel$^{4}$}

\institute{Universitat Pompeu Fabra, Barcelona\\
Corresponding author: \textit{boris.bellalta@upf.edu}
\and
Eindhoven University of Technology, Eindhoven
\and
Hamilton Institute, Maynooth\\
\and
Halmstad University, Halmstad}

\maketitle

\begin{abstract}
This book chapter introduces the use of Continuous Time Markov Networks (CTMN) to analytically capture the operation of Carrier Sense Multiple Access with Collision Avoidance (CSMA/CA) networks. It is of tutorial nature, and it aims to be an introduction on this topic, providing a clear and easy-to-follow description. To illustrate how CTMN can be used, we introduce a set of representative and cutting-edge scenarios, such as Vehicular Ad-hoc Networks (VANETs), Power Line Communication networks and multiple overlapping Wireless Local Area Networks (WLANs). For each scenario, we describe the specific CTMN, obtain its stationary distribution and compute the throughput achieved by each node in the network. Taking the per-node throughput as reference, we discuss how the complex interactions between nodes using CSMA/CA have an impact on system performance. 
\end{abstract}


\section{Introduction}


The presence of devices that use the license-exempt spectrum to communicate is increasing everyday. Those devices range from that of personal and multimedia use (including smart-phones, laptops and storage units, among others) to environment-interactive ones (such as sensors, that gather environmental data, and actuators, that apply a certain action based on given inputs). In between, there is also a plethora of heterogeneous mobile objects, such as vehicles and robots.  

Most of those wireless devices access the channel to transmit data using CSMA/CA, as it offers a good tradeoff between performance and simplicity of implementation. However, when those devices are nearby placed and share the same spectrum band, the use of CSMA/CA creates some complex interactions between their operation, which may also affect their performance. Moreover, those interactions might happen between devices belonging to the same or different networks. For instance, in the former, interactions may occur among two-hop neighbors in a multi-hop network \cite{laufercapacity}, while in the latter, there may be an interplay between multiple co-located WLANs belonging to different owners \cite{herzen2013distributed}.


Analytical models help in improving our understanding of those interactions and allow us to evaluate their impact on system performance. This knowledge should yield to the design of more adequate settings (i.e., parameter configurations) and the development of new mechanisms able to ameliorate the negative effects of such interactions. In this book chapter, we show that CTMN models are able to capture those interactions and provide accurate predictions of their effect on system performance.



To illustrate how CTMN can be used to model the coupled operation between nodes using CSMA/CA, we consider three representative scenarios. For each scenario, we describe the analytical model that captures the behavior of the network, obtain the stationary distribution of the CTMN and compute the per-node throughput. As we discuss in this tutorial, the number of states of the CTMN depends on the number of nodes, their location, the spectrum band they use, and the channel characteristics of the scenario under study. Moreover, the transitions between states are based on the CSMA/CA parameters, such as the backoff-related settings, the packet size and the transmission rate. 
 

This book chapter is structured as follows. First, we present the basic operation of CSMA/CA networks, detailing all assumptions considered in this work. Then, we briefly introduce the required background to understand how the dynamics of CSMA/CA networks can be modelled using CTMN. After that, we model three representative scenarios as use-cases. Finally, we conclude the tutorial summarizing the lessons learned.



\section{Related Work}\label{Sec:RelatedWork}

The use of CTMN models for the analysis of CSMA/CA networks was originally developed in \cite{boorstyn1987throughput} and further extended in the context of IEEE 802.11 networks in \cite{wang2005throughput,durvy2006packing,nardelli2012closed,laufercapacity}, among others. Although the modeling of the IEEE 802.11 backoff mechanism is less detailed than in the work of Bianchi \cite{bianchi2000performance}, it offers greater versatility in modeling a broad range of topologies. Moreover, experimental results in \cite{nardelli2012closed,liew2010back} demonstrate that CTMN models, while idealized, provide remarkably accurate throughput estimates for actual IEEE 802.11 systems.

Boorstyn et al. \cite{boorstyn1987throughput} introduce the use of CTMN models to analyze the throughput of multi-hop CSMA/CA networks. They apply these models to study several network topologies, including a simple chain, a star and a ring network. In \cite{wang2005throughput}, Wang et al., extend the work done in \cite{boorstyn1987throughput} by considering also the fairness between the throughput achieved by each node. Moreover, they connect the parameters of the CTMN with the ones defined by the IEEE 802.11 standard, such as the contention window or the use of RTS/CTS frames. They also provide several approximations with the goal of reducing the model complexity by using local information only. In \cite{durvy2006packing}, Durvy et al., also use CTMN models to characterize the behavior of wireless CSMA/CA networks and explore their spatial reuse gain. Nardelly et al. \cite{nardelli2012closed} extend previous models to specifically consider the negative effect of collisions and hidden terminals. They evaluate 
several multi-hop topologies and compare the results with experimental data, showing that CTMN models can provide very accurate results. In \cite{liew2010back}, Liew et al. also validate the accuracy of CTMN to model CSMA networks using both simulations and experimental data. Besides, they introduce a simple but accurate technique to compute the throughput of each node based on identifying the maximal independent sets of transmitting nodes. Recently, Laufer et al. \cite{laufercapacity} have extended such CTMN models to support non-saturated nodes. Finally, the CTMN model presented in \cite{laufercapacity} is used in \cite{bellaltaperformance} to evaluate the performance of a vehicular video surveillance system.


\section{CSMA/CA Networks}

In this section, we describe the node characteristics and the operation of the considered CSMA/CA protocol. 

\subsection{CSMA/CA protocol}

The MAC protocol defines the rules used by nodes to transmit packets to the channel. The basic principle of CSMA is to listen to the channel before transmitting a packet. In case the channel is detected busy, the transmitter defers its transmission until the channel is sensed idle again. To avoid that all nodes with at least a packet pending for transmission collide as soon as the channel is sensed idle, a collision avoidance (CA) mechanism is introduced. In our case of interest, the CA mechanism is known as the backoff procedure, and it is basically a timer. 
    
The operation of the backoff timer considered in this work is as follows. The backoff timer is initialized to a random value every time the node starts a new transmission attempt. Then, while the channel remains idle, the backoff timer is decreased until it reaches zero, time at which the node transmits the packet to the channel. In case the channel is detected busy before the backoff timer has expired, the node defers the transmission and pauses the backoff timer until the channel becomes idle again.    
    

We also assume that the backoff countdown is continuous in time, and that every time it is initialized, an exponentially distributed random value is selected, with an average duration of $E[B]$ seconds.  Therefore, the attempt rate for every node is equal to $\lambda=E[B]^{-1}$, and the potential\footnote{Potential since the node may be transmitting or overhearing.} activation epochs (i.e., when a node starts a transmission) occur as a Poisson process with rate $\lambda$.

\subsection{Node characteristics}

\begin{figure}[h!!!!]
    \centering    
    \psfrag{t}[][][1.0]{t}
    \psfrag{A}[][][1.0]{A}
    \psfrag{B}[][][1.0]{B}
    \psfrag{C}[][][1.0]{C}
    \psfrag{CR}[][][1.0]{Data Comm. range}
    \psfrag{CSR}[][][1.0]{Carrier Sense range}
    \psfrag{TA}[][][1.0]{$T_A$}
    \psfrag{TB}[][][1.0]{$T_B$}
    \psfrag{BA}[][][1.0]{$B_{A}$}
    \psfrag{BB}[][][1.0]{$B_{B}$}        
    \epsfig{file=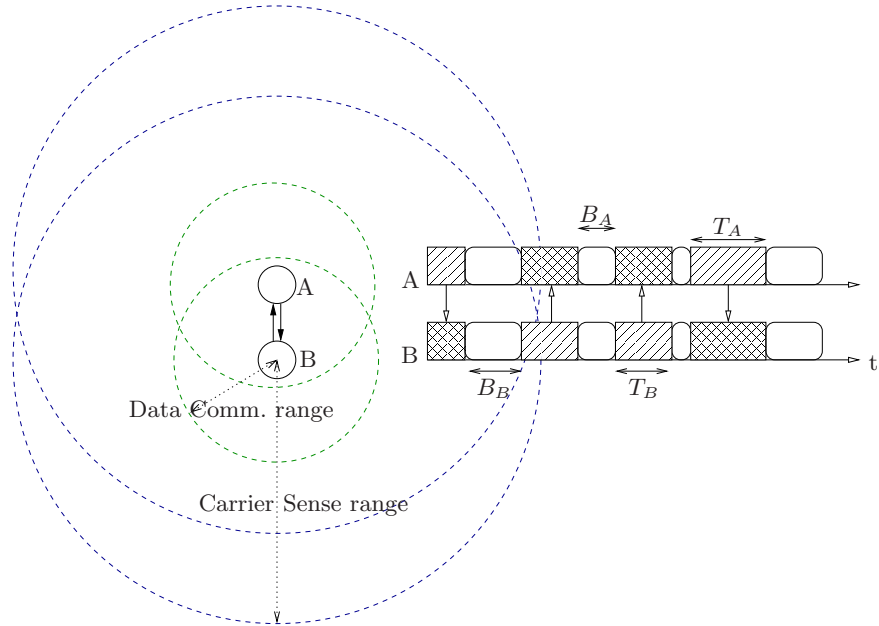,scale=0.5}
    \caption{Two nodes exchanging data using the basic CSMA/CA protocol. $B_i$ indicates the duration of a single backoff instance, and $T_i$ the duration of a packet transmission from node $i$. }
    \label{Fig:CSMAnode}
\end{figure}

Each node implements the CSMA/CA protocol previously described. A node detects the channel busy if the energy level in the channel is equal or higher than the Carrier Sense threshold. However, it will only be able to recover the transmitted data if the energy level has an energy equal or higher than the Data Communication threshold. Based on both thresholds, the carrier sense and data communication ranges are defined. In Figure~\ref{Fig:CSMAnode}, we can see two nodes communicating. The {green} circles represent their data communication ranges, and the {blue} ones their carrier sense ranges.

In this work, we assume that the carrier sense range is at least two times larger than the data communication range. We also assume that the propagation delay between two nodes inside the carrier sense of each other is negligible.  

The Data Communication threshold depends on the transmission rate. It is defined as the required received signal level to guarantee a certain packet error probability when a modulation and a coding rate are employed. Here, we consider that all nodes use a single transmission rate, $R$, and we assume that the considered data communication threshold guarantees an error free transmission regardless the packet size.

All nodes are assumed to be saturated. It means that all nodes have always packets waiting for transmission, and therefore, after transmitting one packet {successfully,} they {will try to} start the transmission of the {next} one. All nodes transmit packets of random size $L$, where $L$ is a random variable exponentially distributed, with average $E[L]$. Therefore, the duration of a packet transmission, $T=L/R$, is also a random variable exponentially distributed, with parameter $\mu={1}/{T}={R}/{L}$. 

Finally, we assume that in case of collision, the affected nodes are able to capture the packet received with a higher energy level.

\subsection{Spatially Distributed CSMA Networks}

Using CSMA/CA, the operation of multiple nodes is coupled if they are inside the carrier sense range of the other, and use the same channel. This happens because as soon as they detect a transmission from the other node, they stop their backoff timer and wait until the channel becomes free to restart it again.

We refer to single-hop networks if all nodes are inside the data communication range of all the other nodes (i.e., any node can transmit packets to the desired destination directly). Otherwise, packets directed to destinations that are outside the data communication range of the transmitter have to be forwarded by intermediary nodes towards it. In this case, we have a multi-hop network. Note that in case of multi-hop networks, nodes that can not communicate directly can also interact between them due to the larger carrier sense range compared with the data communication range, which may cause an undesirable performance loss. 

Finally, nodes from two independent networks can also interact if they operate in the same band and are inside the carrier sense range of each other. In this case, we have a coexistence problem, since both networks see their performance negatively affected.

\subsection{Implications}

Previous assumptions and considerations have the following implications:

\begin{itemize}
    \item \textbf{Collision-free operation between any two nodes that can hear each other}: Given the assumptions of a continuous random backoff and the negligible propagation delay, the probability of packet collisions  between two nodes that can hear each other is negligible.
    \item \textbf{Collisions with hidden nodes and capture effect}: Since the carrier sense range is assumed to be at least two times larger than the data communication range, all nodes that are able to transmit data packets to a target node are able to listen any on-going transmission directed to it and therefore, they can defer their backoff countdown accordingly. However, we consider that the capture effect allows a given receiver to decode the packet directed to it in case of a collision between hidden nodes. Collisions of that nature can happen when transmissions of nodes outside each other's carrier sense range overlap and one of the intended receivers is located inside the intersection of those transmitters' carrier sense ranges.
\end{itemize}


     
\section{Continuous Time Markov Network Models}

In this section we introduce the Continuous Time Markov Network model, which is a stylized Markovian model of $N$~nodes sharing a wireless medium. 

\subsection{Markovian model: the state space and transitions}     
       
Define $\Omega$ as the collection of all feasible \textit{network states}, i.e.~all subsets of the $N$~nodes that can transmit simultaneously, and let $S_t \in \Omega$ be the network state at time $t$. 
We allow for heterogeneous backoff and transmitting rates, i.e.~we assume that at node $i$ the backoff rate is $\lambda_i = E[B_i]^{-1}$ and the transmitting rate is $\mu_i = E[T_i]^{-1}= R_i / E[L_i]$. Then, the transition rates between two network states $s,s' \in \Omega$ are
\begin{equation}
q(s,s')=
\begin{cases}
\lambda_i & \text{ if } s'=s \cup \{ i \} \in \Omega,\\
\mu_i & \text{ if } s'=s \setminus \{ i \},\\
0 & \text{ otherwise}.
\end{cases}
\end{equation}

\subsection{Detailed balanced and product-form stationary distribution}
The process $(S_t)_{t\geq 0}$ has been proven to be a time-reversible continuous-time Markov process in \cite{kelly1979reversibility}. Therefore detailed balance applies and the stationary distribution $\{\pi_s \}_{s \in \Omega}$ can be expressed as in a product-form. Indeed, the detailed balanced relation for two adjacent feasible network states, $s$ and $s \cup \{ i \} \in \Omega$, is

\begin{align}\label{Eq:detailedbalance}
\frac{\pi_{s}}{\pi_{s \cup \{ i \}}} = \frac{\lambda_i}{\mu_i}=\frac{E[B_i]}{E[T_i]}.
\end{align}
For conciseness, we denote $\theta_i:= \lambda_i / \mu_i$. Relation~\eqref{Eq:detailedbalance} gives that for any $s \in \Omega$
\begin{align}
\pi_{s} = \pi_{\emptyset} \cdot \prod_{i \in s} \theta_i,
\end{align}
which, along with the normalizing condition $\sum_{s \in \Omega} \pi_s = 1$, implies that
\begin{align}\label{Eq:product_form}
\pi_{\emptyset} = \frac{1}{\sum_{s \in \Omega} \prod_{i \in s} \theta_i} \quad \text{ and } \quad \pi_{s} = \frac{\prod_{i \in s} \theta_i}{\sum_{s \in \Omega} \prod_{i \in s} \theta_i}, \quad s \in \Omega.
\end{align}
Since the process $(S_t)_{t\geq 0}$ is irreducible and positive recurrent, it follows from classical Markov process results that the stationary distribution $\pi_{s}$ for $s \in \Omega$ is equal to the long-run fraction of time the system spends in the network state $s$. 

Let $x_i$ be the throughput of node $i$, which is computed as follows:

\begin{align}\label{Eq:Throughput}
x_i= \frac{E[L_i]}{E[T_i]}\left(\sum_{s \in \Omega : i \in s}{\pi_s}\right).
\end{align}

\subsection{Insensitivity}

It turns out that for the considered model the stationary distribution $\pi$ (and thus any analytic performance measure linked to it, such as the throughput) is insensitive to the distributions of backoff countdowns and transmission times, in the sense that it depends on these only through the ratios $\theta_i$ of their means. The proof of the insensitivity result can be found in \cite{van2010insensitivity,liew2010back}. The insensitivity property is crucial since the actual behaviour of a network may not be in accordance with backoff and transmission times exponentially distributed.    

                   
                                                            
\section{Scenarios}

We consider three different scenarios to illustrate that CTMN are a suitable tool to model the interactions between different nodes using CSMA/CA to share the medium. In detail, the considered scenarios are:

\begin{itemize}
    \item Vehicular Ad-Hoc networks.
    \item Power Line Communication networks.
    \item Multiple overlapping WLANs supporting channel bonding.
\end{itemize}

All the throughput values plotted in this book chapter have been obtained analytically (i.e., from (\ref{Eq:Throughput})). In case the reader is interested in the accuracy of the model compared with simulation and experimental results, please refer to the papers included in the Related Work section (Section \ref{Sec:RelatedWork}).


\subsection{Vehicular Networks}

Vehicular networks, where vehicles communicate between them, as well as with APs (called Road Service Units, RSUs), are one of the most challenging scenarios for achieving an efficient communication due to the high mobility and rapid topology changes. A general overview of the applications and main challenges of vehicular networks can be found in \cite{hartenstein2008tutorial}. 

\subsubsection{Description of the scenario}            

Overtaking on rural roads often becomes dangerous when oncoming traffic is detected by the driver too late or its speed is underestimated. Recently proposed cooperative overtaking assistance systems, which are based on Vehicular Ad hoc NETworks (VANETs), rely on real-time video transmission. In this case, a video stream captured by a camera installed at the windshield of a vehicle is compressed and broadcast to any vehicles driving behind it, where it is displayed to the driver. Further details regarding this scenario can be found in \cite{vinel2012overtaking}.

\begin{figure}[h!!]
    \centering    
    \psfrag{A}[][][0.80]{A}
    \psfrag{B}[][][0.80]{B}
    \psfrag{C}[][][0.80]{C}
    \psfrag{D}[][][0.80]{D}
    \psfrag{E}[][][0.80]{E}
    \subfigure[Cars in position 1]{\epsfig{file=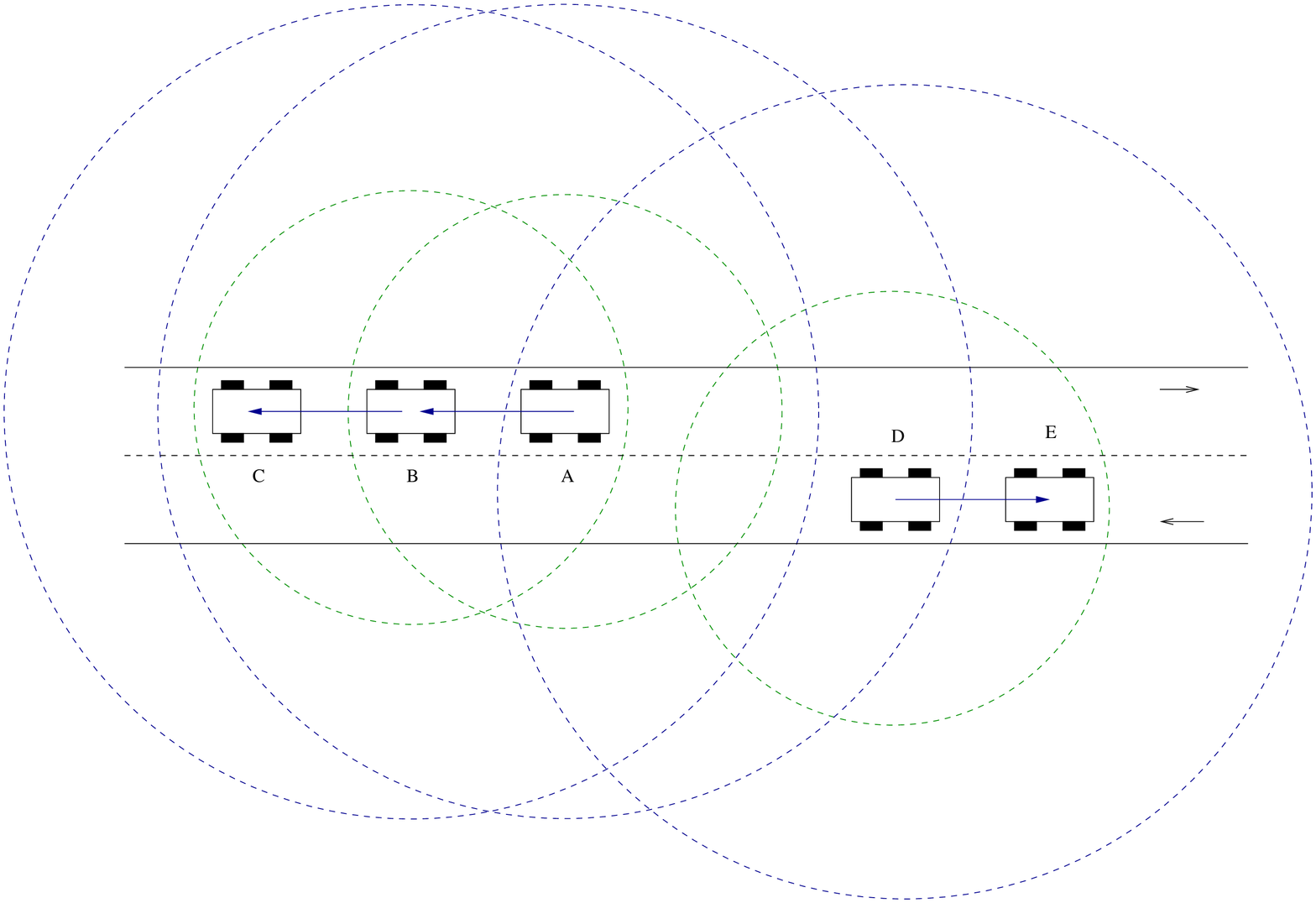,scale=0.35} \label{Fig:case3A}}    
    \subfigure[Cars in position 2]{\epsfig{file=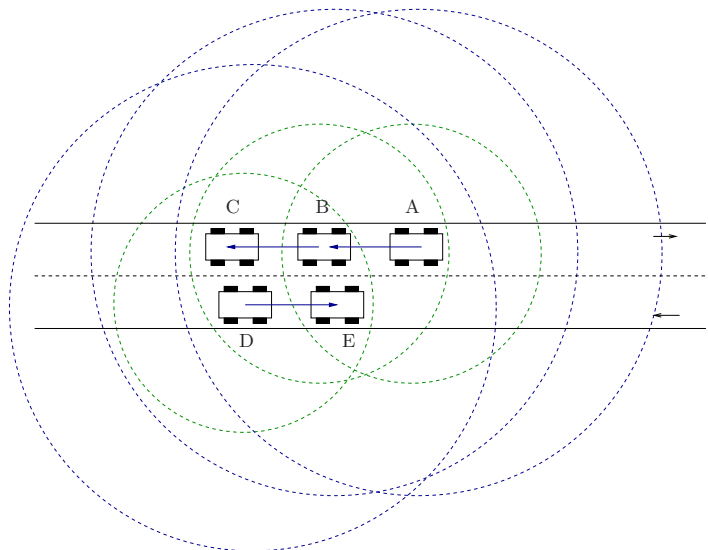,scale=0.35} \label{Fig:case3B}}
    \caption{Two group of cars moving in opposite directions are approaching. In this scenario, the leading car is transmitting a video flow to the cars following it.}
    \label{Fig:case3}
\end{figure}

An example of this scenario formed by two groups of cars is depicted in Figure~\ref{Fig:case3}. The first group formed by cars A, B and C moves from the left to the right, and the second group, formed by cars D and E, moves in the opposite direction. In detail, car A sends data to car B, car B sends data to car C, and car D sends data to car E. Note that cars C and E do not transmit packets. In this case, we can consider that the exchanged data corresponds to a video stream generated by the platoon's leading car. To illustrate the effect of mobility on the network performance, two different positions of the cars are considered. The first position represents the case where both group of cars are approaching each other. The second position represents the case where the two group of cars are side by side.

\subsubsection{Model}

\begin{figure}[h!!!]
    \centering 
    \psfrag{0}[][][1.0]{$\emptyset$}
    \psfrag{A}[][][1.0]{A}
    \psfrag{B}[][][1.0]{B}
    \psfrag{C}[][][1.0]{C}
    \psfrag{D}[][][1.0]{D}
    
    \psfrag{BD}[][][1.0]{BD}
    \psfrag{lAuA}[][][1.0]{$\lambda_A,\mu_A$}
    \psfrag{lBuB}[][][1.0]{$\lambda_B,\mu_B$}
    \psfrag{lCuC}[][][1.0]{$\lambda_D,\mu_D$}
    \psfrag{lDuD}[][][1.0]{$\lambda_D,\mu_D$}
    
    \subfigure[Cars in position 1]{\epsfig{file=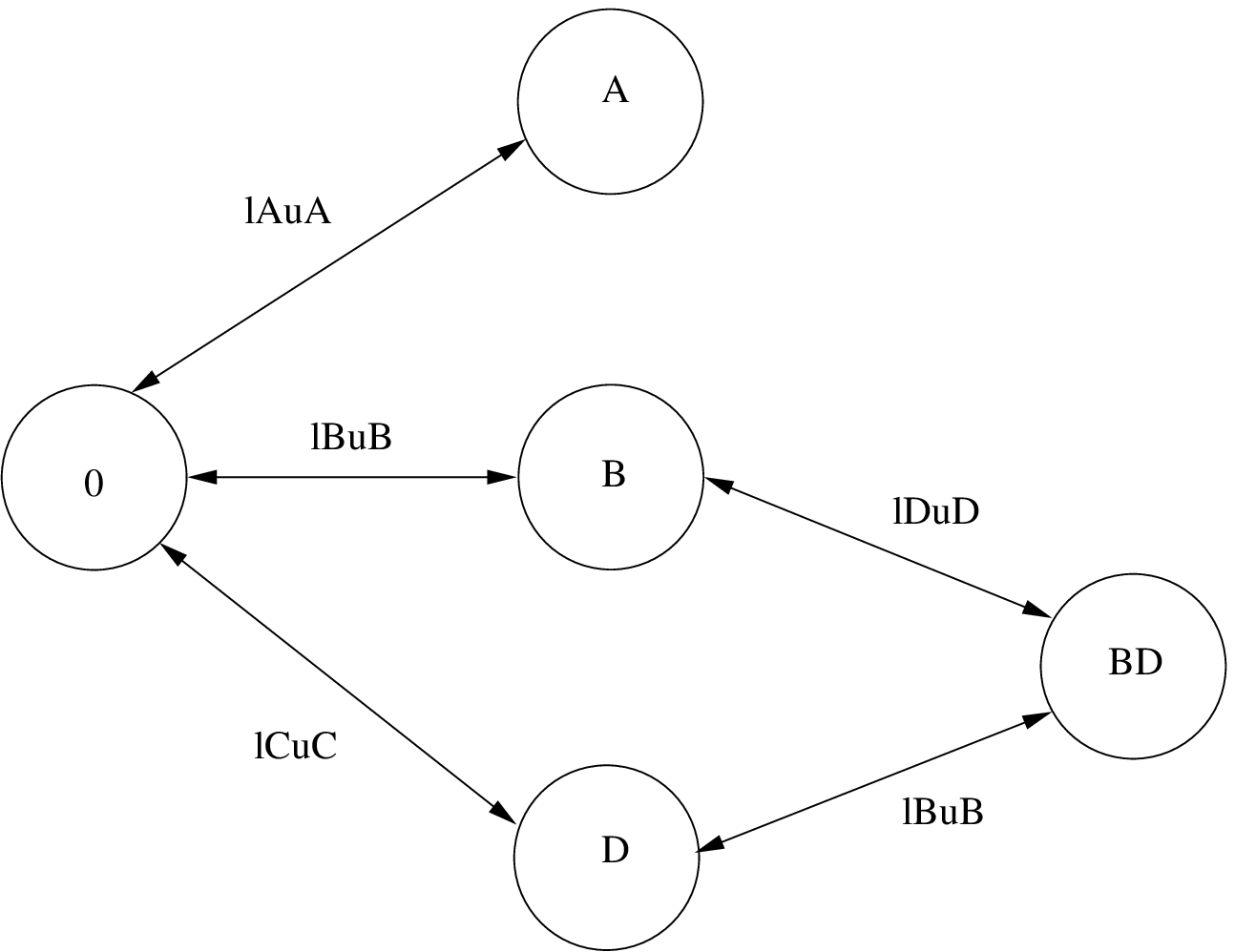,scale=0.5}\label{Fig:MC_case2A}} \quad \quad \quad
    \subfigure[Cars in position 2]{\epsfig{file=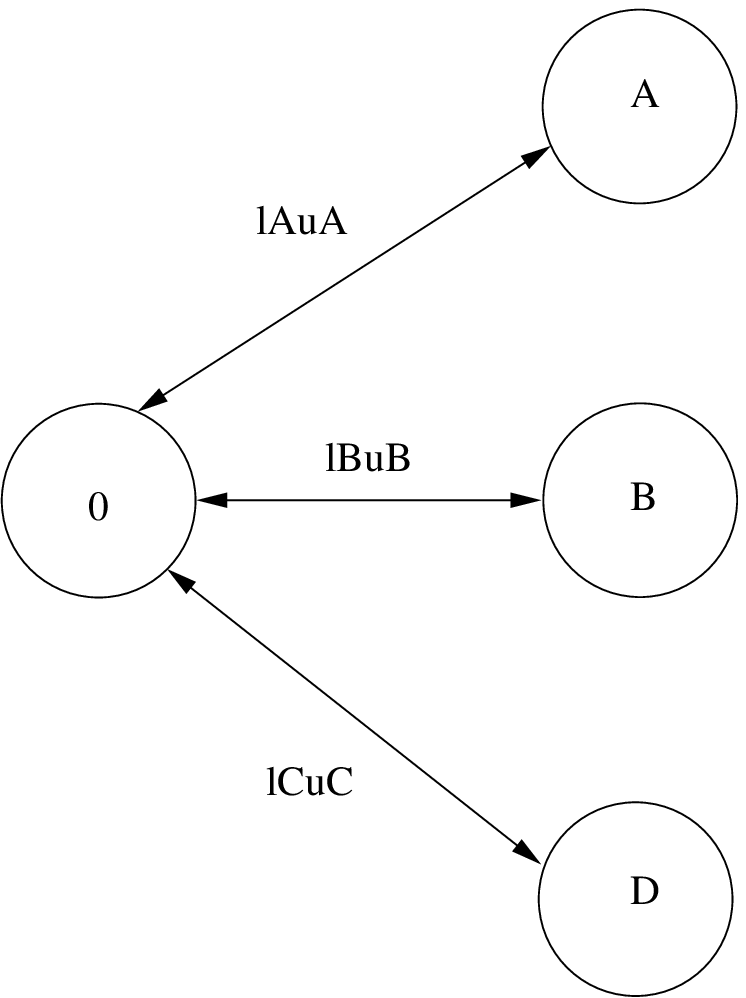,scale=0.5}\label{Fig:MC_case2B}}
    
    \caption{Markov network for the vehicular scenario.}
    \label{Fig:MC_case2}
\end{figure}

In Figures \ref{Fig:MC_case2A} and \ref{Fig:MC_case2B}, we show the CTMNs that capture the feasible states of the vehicular networks depicted in Figure~\ref{Fig:case3A} (position 1) and Figure~ \ref{Fig:case3B} (position 2), respectively.

Considering first the case in which the cars are in position 1, at the equilibrium, the mean fraction of time each state is active (i.e., the stationary distribution) is given by:

\begin{align}
    \pi_{\emptyset}&=\frac{1}{1+\theta_A+\theta_B+\theta_D+\theta_B\theta_D}~~~~~~~
    \pi_{A}=\frac{\theta_A}{1+\theta_A+\theta_B+\theta_D+\theta_B\theta_D} \nonumber\\
    \pi_{B}&=\frac{\theta_B}{1+\theta_A+\theta_B+\theta_D+\theta_B\theta_D} ~~~~~~~
    \pi_{D}=\frac{\theta_D}{1+\theta_A+\theta_B+\theta_D+\theta_B\theta_D} \nonumber\\
    \pi_{BD}&=\frac{\theta_B\theta_D}{1+\theta_A+\theta_B+\theta_D+\theta_B\theta_D}  \nonumber       
\end{align}
with $\theta_A=\lambda_A E[T_A]$, $\theta_B=\lambda_B E[T_B]$, and $\theta_D=\lambda_D E[T_D]$.

However, since all cars have the same $\lambda$ and $\mu$ parameters, we have that $\theta=\lambda/\mu$. Then,

\begin{align}
    \pi_{\emptyset}&=\frac{1}{1+3\theta+\theta^2} ~~~~~~~
    \pi_{A}=\frac{\theta}{1+3\theta+\theta^2}   ~~~~~~~
    \pi_{B}=\frac{\theta}{1+3\theta+\theta^2} \nonumber\\
    \pi_{D}&=\frac{\theta}{1+3\theta+\theta^2}  ~~~~~~~
    \pi_{BD}=\frac{\theta^2}{1+3\theta+\theta^2}  \nonumber       
\end{align}

Finally, the throughput achieved by each car is given by:

\begin{align}
    x_{A}=\frac{E[L_A]}{E[T_A]}\left(\pi_A\right) ~~~~~~~~ 
    x_{B}=\frac{E[L_B]}{E[T_B]}\left(\pi_B + \pi_{BD} \right) ~~~~~~~~ 
    x_{D}=\frac{E[L_D]}{E[T_D]}\left(\pi_D + \pi_{BD} \right) \nonumber
\end{align}

As it can be seen, the throughput achieved by cars B and D is higher than the throughput achieved by car A. This situation might harm the quality of the video sent by car A, causing packet losses due to buffer overflow. 

Similarly, when cars are in position 2, all of them are in the coverage area of the others. Then, the mean fraction of time each state is active is given by:

\begin{align}
    \pi_{\emptyset}=\frac{1}{1+3\theta} ~~~~~~~~ 
    \pi_{A}=\frac{\theta}{1+3\theta} ~~~~~~~~ 
    \pi_{B}=\frac{\theta}{1+3\theta} ~~~~~~~~ 
    \pi_{D}=\frac{\theta}{1+3\theta} \nonumber
\end{align}

which results in the same throughput for each car:

\begin{align}
    x_{A}=\frac{E[L_A]}{E[T_A]}\pi_A~~~~~~~~ x_{B}=\frac{E[L_B]}{E[T_B]}\pi_B~~~~~~~~ x_{D}=\frac{E[L_D]}{E[T_D]}\pi_D \nonumber
\end{align}
                                     
\subsubsection{Numerical Results and Discussion}

\begin{figure}[h!!!]
    \centering
    \epsfig{file=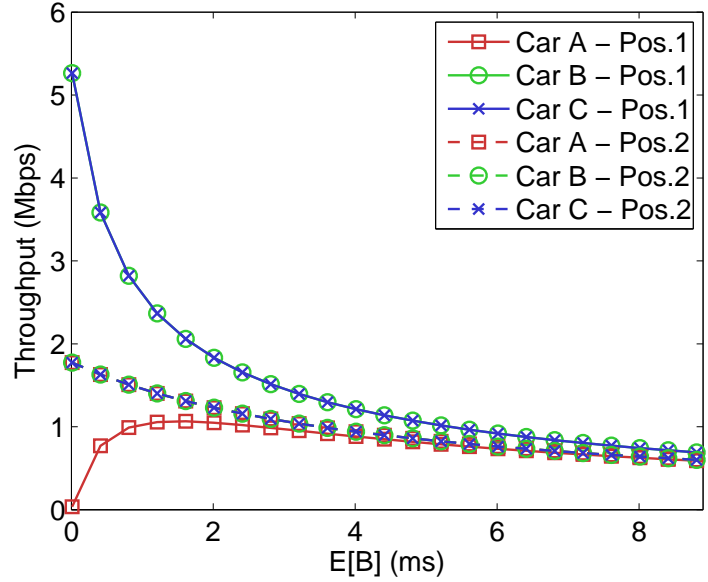,scale=0.75}
    \caption{Throughput achieved by each car.}
    \label{Fig:S_vehicular}
\end{figure}

In Figure~\ref{Fig:S_vehicular}, we plot the throughput achieved by each car versus the expected backoff duration, $E[B]$. For all nodes, the expected duration of a packet transmission is $E[T]=3$ ms, and the average packet size is $E[L]=8000$ bits.

It can be observed that, when cars are in position 1, for very low $E[B]$ values, car A suffers from complete starvation. Higher $E[B]$ values increase the chances for car A to transmit, which increases its throughput. However, the throughput of car A will be always below the throughput achieved by cars B and D. When cars move to position 2, since all of them are inside the coverage area of the others, all vehicles achieve the same throughput. This example shows how mobility may severely affect the network performance.

            

\subsection{Multi-hop Power Line Communication Networks}

Power Line Communication (PLC) networks are formed by devices interconnected using electrical wires. Despite being classified as wired instead of wireless networks, PLC networking has several factors in common with wireless connectivity. PLC channels, as well as in wireless networks, are affected by propagation impairments and effects like hidden and exposed terminals \cite{ferreira1999power}. These characteristics make the use of traditional medium access protocols for wired networks not suitable to PLC. In contrast, current PLC standards, such as Homeplug \cite{HomeplugStd} and IEEE 1901 \cite{IEEE1901}, use a CSMA/CA approach very similar in nature to that defined in IEEE 802.11 Distributed Coordination Function (DCF). 

The use of PLC devices is expected to grow in the years to come due to the recent availability of affordable and off-the-self devices. However, further research on higher than the physical layer needs to be carried out to fully demonstrate the capabilities and limitations of this technology. For instance, the evaluation of multi-hop PLC networks is still an open research area that imposes several challenges. The analytical framework presented here allows us to give a step further in this evaluation. Observe that, due to propagation impairments, multi-hop communication may be needed in certain PLC topologies, where the path between two communication pairs makes a direct communication not viable. That is the case of long electrical wires used in grids but also the case of large or signal-propagation-challenging buildings.

\subsubsection{Description of the scenario}

\begin{figure}[h!!!]
    \centering
    
    \psfrag{A}[][][1.0]{A}
    \psfrag{B}[][][1.0]{B}
    \psfrag{C}[][][1.0]{C}
    \psfrag{D}[][][1.0]{D}
    \psfrag{E}[][][1.0]{E}
    \psfrag{Sink}[][][1.0]{Gateway}
    \psfrag{DR}[][][1.0]{Comm. range}
    \psfrag{CSR}[][][1.0]{Carrier Sense range}

    \epsfig{file=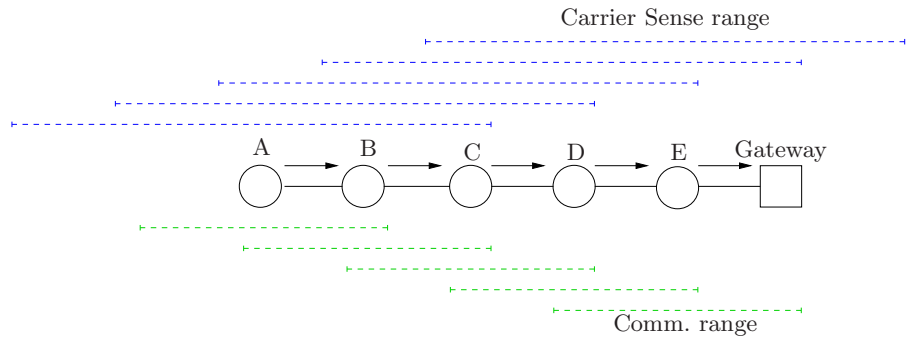,scale=0.55}
    \caption{Multihop PLC scenario.}
    \label{Fig:case2}
\end{figure}

Consider the multi-hop power line network depicted in Figure~\ref{Fig:case2}, where effective connectivity (not actual wiring) and carrier sense ranges are depicted. This scenario is representative of a large building where nodes equipped with PLC modules connected to the mains run data-intensive applications. It also applies to other deployments, such as outdoors video surveillance with devices connected to urban furniture with access to the electrical grid, such as lampposts. In both cases, we assume nodes are not able to reach the central unit (point at which the data is processed, stored or sent to a server in the cloud) directly. In contrast, nodes must send their traffic sequentially using the other nodes in the chain as relays. 

We assume here that nodes inside the carrier sense range are able to decode the delimiter of a frame \cite{HomeplugStd,IEEE1901} but do not receive data correctly. Thus, they defer their transmissions when overhearing one and two-hop neighboring nodes. For instance, when node C transmits, all the other four nodes will sense the channel busy and will defer their backoff. Node C will only be allowed to transmit when all the other four nodes are idle, which, as we will see, severely affects its performance. On the contrary, nodes A and D can transmit simultaneously. The same applies for nodes A and E, as well as B and E. 

We do not consider here sophisticated features of the standard that can influence the results, such as aggregation, frame bursting, contention free channel access, arbitration and flow control \cite{HomeplugStd,IEEE1901}.

\subsubsection{Model}

\begin{figure}[h!!!]
    \centering
    \psfrag{0}[][][1.0]{$\emptyset$}
    \psfrag{A}[][][1.0]{A}
    \psfrag{B}[][][1.0]{B}
    \psfrag{C}[][][1.0]{C}
    \psfrag{D}[][][1.0]{D}
    \psfrag{E}[][][1.0]{E}
    \psfrag{AD}[][][1.0]{AD}
    \psfrag{AE}[][][1.0]{AE}
    \psfrag{BE}[][][1.0]{BE}
    \psfrag{lAuA}[][][1.0]{$\lambda_A,\mu_A$}
    \psfrag{lBuB}[][][1.0]{$\lambda_B,\mu_B$}
    \psfrag{lCuC}[][][1.0]{$\lambda_C,\mu_C$}
    \psfrag{lDuD}[][][1.0]{$\lambda_D,\mu_D$}
    \psfrag{lEuE}[][][1.0]{$\lambda_E,\mu_E$}
    \psfrag{lAuD}[][][1.0]{$\lambda_A,\mu_D$}
    \psfrag{lAuE}[][][1.0]{$\lambda_A,\mu_E$}
    \psfrag{lBuE}[][][1.0]{$\lambda_B,\mu_E$}
    \psfrag{lDuA}[][][1.0]{$\lambda_A,\mu_A$}
    \psfrag{lEuA}[][][1.0]{$\lambda_E,\mu_A$}
    \psfrag{lEuB}[][][1.0]{$\lambda_E,\mu_B$}

    \psfrag{lAuA}[][][1.0]{}
    \psfrag{lBuB}[][][1.0]{}
    \psfrag{lCuC}[][][1.0]{}
    \psfrag{lDuD}[][][1.0]{}
    \psfrag{lEuE}[][][1.0]{}
    \psfrag{lAuD}[][][1.0]{}
    \psfrag{lAuE}[][][1.0]{}
    \psfrag{lBuE}[][][1.0]{}
    \psfrag{lDuA}[][][1.0]{}
    \psfrag{lEuA}[][][1.0]{}
    \psfrag{lEuB}[][][1.0]{}

    \epsfig{file=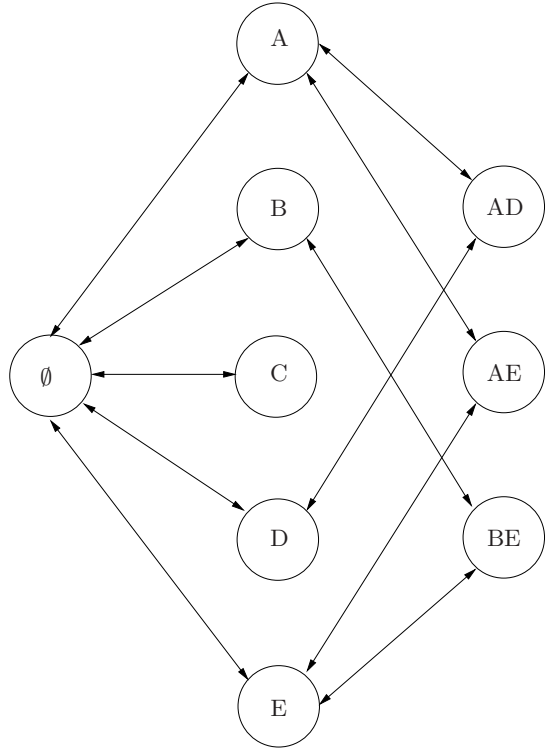,scale=0.55}
    \caption{CTMN representing the PLC scenario. We have not represented the transition rates for the sake of simplicity of illustration.}
    \label{Fig:MC_case3}
\end{figure}

The feasible states of the considered PLC network can be represented using the CTMN shown in Figure~\ref{Fig:MC_case3}, where each state represents a group of nodes that are active simultaneously. For instance, state A means that only node A is transmitting, while state AD means that both nodes A and D are simultaneously transmitting. As we can see, the state space of the CTMN is affected by both the network topology and the carrier sense range.

At the equilibrium, the mean fraction of time each state is active is given by:

\begin{align}
    \pi_{\emptyset}&=\frac{1}{1+\theta_A+\theta_B+\theta_C+\theta_D+\theta_E + \theta_A\theta_D+ \theta_A\theta_E+ \theta_B\theta_E} \nonumber\\
    \pi_{A}&=\frac{\theta_A}{1+\theta_A+\theta_B+\theta_C+\theta_D+\theta_E + \theta_A\theta_D+ \theta_A\theta_E+ \theta_B\theta_E} \nonumber\\
    \pi_{B}&=\frac{\theta_B}{1+\theta_A+\theta_B+\theta_C+\theta_D+\theta_E + \theta_A\theta_D+ \theta_A\theta_E+ \theta_B\theta_E} \nonumber\\
    \pi_{C}&=\frac{\theta_C}{1+\theta_A+\theta_B+\theta_C+\theta_D+\theta_E + \theta_A\theta_D+ \theta_A\theta_E+ \theta_B\theta_E} \nonumber\\
    \pi_{D}&=\frac{\theta_D}{1+\theta_A+\theta_B+\theta_C+\theta_D+\theta_E + \theta_A\theta_D+ \theta_A\theta_E+ \theta_B\theta_E} \nonumber\\
    \pi_{E}&=\frac{\theta_E}{1+\theta_A+\theta_B+\theta_C+\theta_D+\theta_E + \theta_A\theta_D+ \theta_A\theta_E+ \theta_B\theta_E} \nonumber\\
    \pi_{AD}&=\frac{\theta_A\theta_D}{1+\theta_A+\theta_B+\theta_C+\theta_D+\theta_E + \theta_A\theta_D+ \theta_A\theta_E+ \theta_B\theta_E}  \nonumber    \\   
    \pi_{AE}&=\frac{\theta_A\theta_E}{1+\theta_A+\theta_B+\theta_C+\theta_D+\theta_E + \theta_A\theta_D+ \theta_A\theta_E+ \theta_B\theta_E}  \nonumber \\      
    \pi_{BE}&=\frac{\theta_B\theta_E}{1+\theta_A+\theta_B+\theta_C+\theta_D+\theta_E + \theta_A\theta_D+ \theta_A\theta_E+ \theta_B\theta_E},  \nonumber      
\end{align}
with $\theta_A=\lambda_A E[T_A]$, $\theta_B=\lambda_B E[T_B]$, $\theta_C=\lambda_C E[T_C]$, $\theta_D=\lambda_D E[T_D]$, and $\theta_E=\lambda_E E[T_E]$.

From the stationary distribution, we can obtain the throughput achieved by each node:

\begin{align}
    x_{A}&=\frac{E[L_A]}{E[T_A]}\left(\pi_A + \pi_{AD}  + \pi_{AE} \right) \nonumber\\
    x_{B}&=\frac{E[L_B]}{E[T_B]}\left(\pi_B + \pi_{BD} \right) \nonumber\\
    x_{C}&=\frac{E[L_C]}{E[T_C]}\left(\pi_C \right) \nonumber\\
    x_{D}&=\frac{E[L_D]}{E[T_D]}\left(\pi_D + \pi_{AD} \right) \nonumber\\
    x_{E}&=\frac{E[L_E]}{E[T_E]}\left(\pi_E + \pi_{AE} + \pi_{BE} \right) \nonumber   
\end{align}

\subsubsection{Numerical Results and Discussion}

\begin{figure}[h!!!]
    \centering
    \epsfig{file=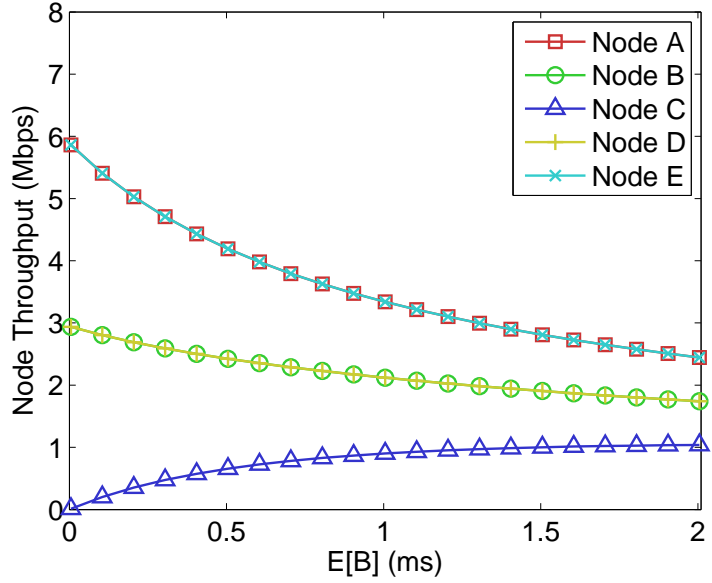,scale=0.75}
    \caption{Throughput achieved by each PLC node.}
    \label{Fig:S_sensors}
\end{figure}

In Figure~\ref{Fig:S_sensors}, we plot the throughput of each node with respect to the $E[B]$ duration. For all nodes, the considered average packet size is $E[L]=12000$ bits, that results in an average packet transmission duration of $E[T]=1359.02$ $\mu s$ using Homeplug 1.0 parameters \cite{HomeplugStd}.

We can observe that nodes A and E achieve the same saturation throughput. As expected, due to the network symmetry, nodes B and D also achieve the same throughput. Clearly, the network bottleneck is node C. Although not completely solving the unfair effect, increasing $E[B]$ at all nodes ameliorates its magnitude.

\subsection{Channel Bonding in WLANs}

Multimedia communications between multimedia devices, such as smart TVs, high definition video and music players, file storage servers, tablets, and laptops is one of the scenarios targeted by next generation WLANs. One of the strategies that can be used to satisfy the performance requirements of those applications in WLANs is channel bonding, which simply consists on the use of wider channels. The use of wider channels in WLANs has been considered recently in the IEEE 802.11n amendment \cite{IEEE80211n} and further expanded in the IEEE 802.11ac amendment \cite{IEEE80211ac}. A wider channel is obtained by grouping several $20$ MHz basic channels.

While the use of wider channels allows faster packet transmissions \cite{Ray2013}, it also increases the chances to overlap with other WLANs. Therefore, it is not obvious whether the resulting performance is improved compared to the single channel case.

\subsubsection{Description of the scenario}    
        
\begin{figure}[h!!!]
    \centering
    \psfrag{A}[][][1.0]{A}
    \psfrag{B}[][][1.0]{B}
    \psfrag{C}[][][1.0]{C}
    \psfrag{D}[][][1.0]{D}
    \psfrag{E}[][][1.0]{E}
    \epsfig{file=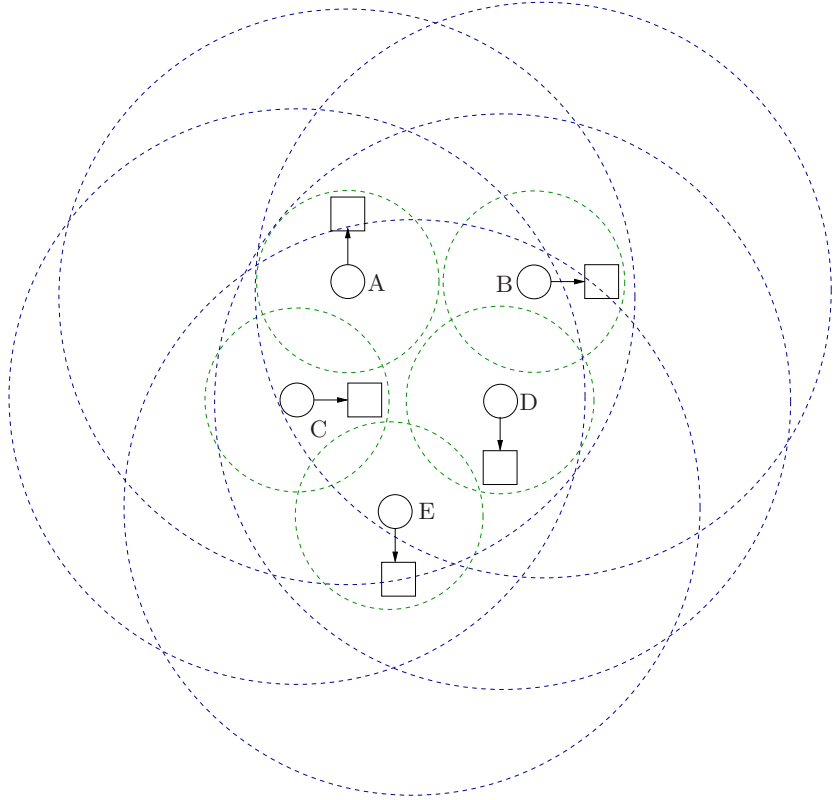,scale=0.45}
    \caption{Five co-located WLANs. All APs are inside the carrier sense area of all others. The channels used by each AP are shown in Figure~\ref{Fig:Channels_Case1}.}
    \label{Fig:case1}
\end{figure}

\begin{figure}[h!!!]
    \centering
    
    \psfrag{A}[][][1.0]{A}
    \psfrag{B}[][][1.0]{B}
    \psfrag{C}[][][1.0]{C}
    \psfrag{D}[][][1.0]{D}
    \psfrag{E}[][][1.0]{E}
    \psfrag{channels}[][][1.0]{channels}
    \psfrag{1}[][][1.0]{1}
    \psfrag{2}[][][1.0]{2}
    \psfrag{3}[][][1.0]{3}
    \psfrag{4}[][][1.0]{4}
    \psfrag{5}[][][1.0]{5}
    \psfrag{6}[][][1.0]{6}
    \psfrag{7}[][][1.0]{7}
    \psfrag{8}[][][0.9]{8}
    \epsfig{file=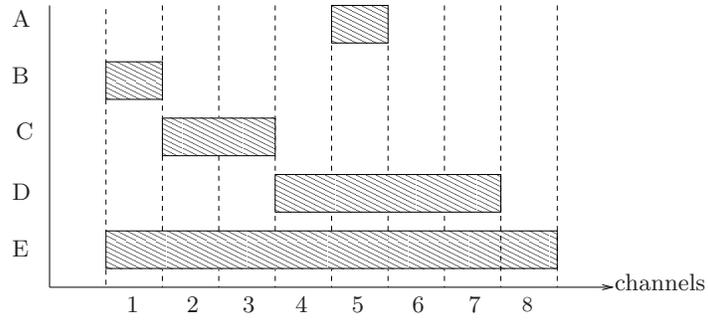,scale=0.5}
    \caption{Channels allocated to each WLAN.}
    \label{Fig:Channels_Case1}
\end{figure}

In Figure~\ref{Fig:case1}, we show a scenario with 5 co-located WLANs. As it is shown in Figure~\ref{Fig:Channels_Case1}, WLANs~A and~B use a single basic channel, WLAN~C uses two basic channels, WLAN~D uses $4$ basic channels and WLAN~E uses $8$ basic channels. We consider that increasing the channel width $c$ times reduces the transmission time of a packet by the same factor $c$, i.e., the mean transmission time of a packet using $c$ basic channels is $E[T_i]/c$, with $E[T_i]$ the transmission time when a single channel is used. In other words, we are not considering the performance loss caused by headers and control frames that are duplicated in every basic channel.    

\subsubsection{Model}

\begin{figure}[h!!!]
    \centering
    \psfrag{0}[][][1.0]{$\emptyset$}    
    \psfrag{A}[][][1.0]{A}
    \psfrag{B}[][][1.0]{B}
    \psfrag{C}[][][1.0]{C}
    \psfrag{D}[][][1.0]{D}
    \psfrag{E}[][][1.0]{E}    
    \psfrag{AB}[][][1.0]{AB}
    \psfrag{AC}[][][1.0]{AC}
    \psfrag{BC}[][][1.0]{BC}
    \psfrag{BD}[][][1.0]{BD}
    \psfrag{CD}[][][1.0]{CD}
    \psfrag{ABC}[][][1.0]{ABC}
    \psfrag{BCD}[][][1.0]{BCD}
    \psfrag{lAuA}[][][0.9]{$\lambda_A,\mu_A$}
    \psfrag{lBuB}[][][0.9]{$\lambda_B,\mu_B$}
    \psfrag{lCuC}[][][0.9]{$\lambda_C,\mu_C$}
    \psfrag{lDuD}[][][0.9]{$\lambda_D,\mu_D$}
    \psfrag{lEuE}[][][0.9]{$\lambda_E,\mu_E$}

    \psfrag{lAuA}[][][0.9]{}
    \psfrag{lBuB}[][][0.9]{}
    \psfrag{lCuC}[][][0.9]{}
    \psfrag{lDuD}[][][0.9]{}
    \psfrag{lEuE}[][][0.9]{}

    \epsfig{file=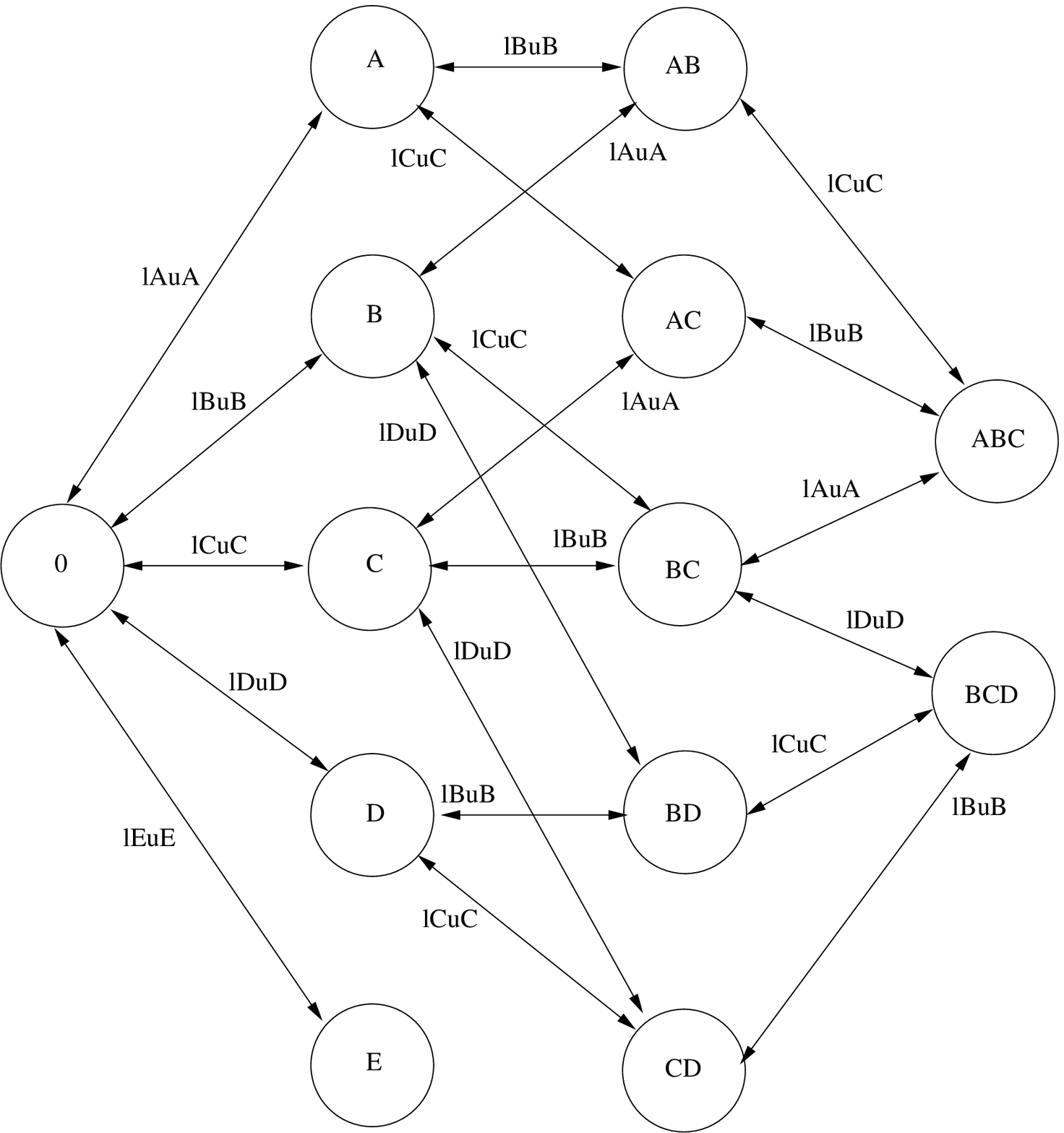,scale=0.55}
    \caption{CTMN that represents the multiple overlapping WLANs scenario. We have not represented the transition rates for the sake of simplicity of illustration.}
    \label{Fig:MC_case1}
\end{figure}

In Figure~\ref{Fig:MC_case1}, we show the CTMN that captures the feasible states that represent the dynamics of the group of co-located WLANs shown in Figure~\ref{Fig:case1}. Note that two WLANs overlap if they share at least one single channel and, therefore, they cannot be active simultaneously.

At the equilibrium, the mean fraction of time each state is active is given by:
\begin{align}
    \pi_{\emptyset}&=\frac{1}{\phi}~~~~~~~\pi_{A}=\frac{\theta_A}{\phi}~~~~~~~ \pi_{B}=\frac{\theta_B}{\phi}~~~~~~~
    \pi_{C}=\frac{\theta_C}{\phi}~~~~~~~   \pi_{D}=\frac{\theta_D}{\phi}~~~~~~~ \pi_{E}=\frac{\theta_E}{\phi}\nonumber\\
    \pi_{AB}&=\frac{\theta_A\theta_B}{\phi}~~~~~~~\pi_{AC}=\frac{\theta_A \theta_C}{\phi}~~~~~~~\pi_{BC}=\frac{\theta_B \theta_C}{\phi}~~~~~~~
    \pi_{BD}=\frac{\theta_B \theta_D}{\phi}\nonumber\\
\pi_{CD}&=\frac{\theta_C \theta_D}{\phi}  ~~~~~~~ \pi_{ABC}=\frac{\theta_A\theta_B\theta_C}{\phi}~~~~~~~\pi_{BCD}=\frac{\theta_B \theta_C\theta_D}{\phi},\nonumber
\end{align}
with $\theta_A=\lambda_A E[T_A]$, $\theta_B=\lambda_B E[T_B]$, $\theta_C=\lambda_C E[T_C]/2$, $\theta_D=\lambda_D E[T_D]/4$, $\theta_E=\lambda_E E[T_E]/8$, and  $\phi=1+\theta_A+\theta_B+\theta_C+\theta_D+\theta_E+\theta_A\theta_B+\theta_A\theta_C+\theta_B\theta_C+\theta_B\theta_D+\theta_C\theta_D+\theta_A\theta_B\theta_C+\theta_B\theta_C\theta_D$.

We can obtain the mean throughput of WLAN $i$ by multiplying the total time each WLAN is active by the WLAN bitrate, $\frac{E[L_i]}{E[T_i]/c}$, i.e.,

\begin{align}
    x_{A}&=(\pi_{A}+\pi_{AB}+\pi_{AC}+\pi_{ABC})\frac{E[L_A]}{E[T_A]} \nonumber\\
    x_{B}&=(\pi_{B}+\pi_{AB}+\pi_{BC}+\pi_{BD}+\pi_{ABC}+\pi_{BCD})\frac{E[L_B]}{E[T_B]} \nonumber\\
    x_{C}&=(\pi_{C}+\pi_{AC}+\pi_{BC}+\pi_{CD}+\pi_{ABC}+\pi_{BCD})\frac{E[L_C]}{E[T_C]/2} \nonumber\\
    x_{D}&=(\pi_{D}+\pi_{BD}+\pi_{CD}+\pi_{BCD})\frac{E[L_D]}{E[T_D]/4} \nonumber\\
    x_{E}&=(\pi_{E})\frac{E[L_E]}{E[T_E]/8} \nonumber  
\end{align}

\subsubsection{Numerical Results and Discussion}

\begin{figure}
    \centering
    \epsfig{file=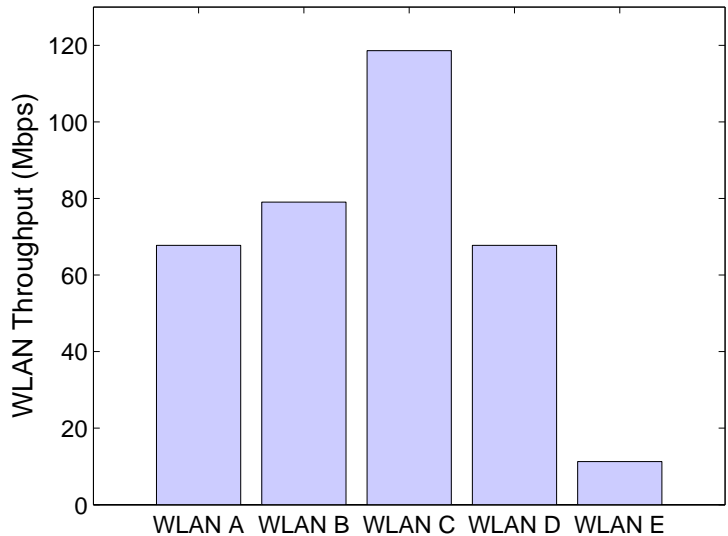,scale=0.75}
    \caption{Throughput achieved by each WLAN.}
    \label{Fig:S_wlans}
\end{figure}

In Figure~\ref{Fig:S_wlans}, we plot the throughput achieved by each WLAN when $E[T]=0.1$ ms, $E[B]=50~\mu$s, and $E[L]=12000$ bits for all WLANs. We observe that WLAN~B achieves a higher throughput than WLAN~A since it has less contenders. Similarly to WLAN~B, WLAN~C only contends with WLAN~E. However, WLAN~C achieves a higher throughput than WLAN~B, since it uses a channel two times wider. WLAN~D achieves the same throughput as WLAN~A in spite of using a channel four times wider. This situation is known as performance anomaly, and was described in \cite{heusse2003performance} for the case in which different nodes use different transmission rates. WLAN~E uses the widest channel compared to the other WLANs. However, it is also the WLAN with more contenders, some of them independent of the others. This situation causes WLAN~E to be inactive for long periods, resulting in the WLAN with the lowest throughput.

\section{Summary}
    
In this book chapter we have shown that Continuous Time Markov Networks can be applied to model CSMA/CA networks. To illustrate how this kind of models can be used, we have considered them to model three different toy scenarios: vehicular ad-hoc networks, PLC networks and multiple overlapping WLANs using channel bonding.

For each scenario, we have described the state space and the transitions between states by drawing its corresponding CTMN. Then, we have computed the stationary distribution of the Markov network and the per-node throughput. After that, we have evaluated the network performance, focusing on describing how the different nodes interact due to the use of CSMA/CA.


\bibliographystyle{unsrt}
\bibliography{TheBib}

\end{document}